\documentclass[galaxies,review,accept,moreauthors]{mdpi} 
\firstpage{1} 
\pubvolume{1}
\issuenum{1}
\articlenumber{0}
\pubyear{2025}
\copyrightyear{2025}
\externaleditor{Oleg Malkov} 
\datereceived{28 February 2025} 
\daterevised{3 April 2025} 
\dateaccepted{14 April 2025} 
\datepublished{22 April 2025} 
\hreflink{https://doi.org/} 

\Title{Intermediate-Mass Mergers: A New Scenario for Several 
FS~CMa
Stars}

\TitleCitation{Intermediate-Mass Mergers: A New Scenario for Several FS CMa Stars}

\Author{Daniela 
Kor\v{c}\'{a}kov\'{a} 
 $^{1,}$*\orcidA{}, 
Nela~Dvo\v{r}\'{a}kov\'{a} $^{1}$,
Iris~Bermejo Lozano $^{1}$, 
Gregg~A.~Wade $^{2,3}$, 
Alicia~Moranchel Basurto $^{1}$, 
Pavel~Kroupa $^{1,4}$, 
Raul~Ortega~Chametla $^{1}$,
Colin~Peter~Folsom $^{5}$ and
Ondrej Juh\'{a}s $^{1}$
}

\AuthorNames{
Daniela Kor\v{c}\'{a}kov\'{a}
Nela~Dvo\v{r}\'{a}kov\'{a},
Iris~Bermejo Lozano, 
Gregg~A.~Wade, 
Alicia~Moranchel Basurto, 
Pavel~Kroupa, 
Raul~Ortega~Chametla and
Colin~Peter~Folsom
Ondrej Juh\'{a}s
}

{
        \AuthorCitation{Kor\v{c}\'{a}kovxax, D.; Dvo\v{r}xaxkovxax, N.; Lozano, I.B.; Wade, G.A.; Basurto, A.M.; Kroupa, P.; Chametla, R.O.; Juhxaxs, C.P.F.}
        
}

\address{%
$^{1}$ \quad {Astronomical Institute, Charles University in Prague, V Hole\v{s}ovi\v{c}k\'{a}ch 2, Praha 8, 18000 Prague
, \linebreak Czech Republic; kor@sirrah.troja.mff.cuni.cz (D.K.); dvorakova.nela@seznam.cz (N.D.); \linebreak irisberloz@gmail.com (I.B.L.);  amoranchel087@gmail.com (A.M.B.); pavel.kroupa@mff.cuni.cz (P.K.); rortegaesfm@gmail.com (R.O.C.); ondrej.juhas12@gmail.com (O.J.) 
} \\
$^{2}$ \quad {Department of Physics, Engineering Physics \& Astronomy, Queen's University, Kingston, ON K7L 3N6, Canada; gregg.wade@rmc.ca} (G.A.W.) \\ 
$^{3}$ \quad {Department of Physics \& Space Science, Royal Military College of Canada, PO Box 17000, Station Forces, Kingston, ON K7K 7B4, Canada}\\
$^{4}$ \quad {Helmholtz-Institut f\"{u}r Strahlen- und Kernphysik, Universitaet Bonn, Nussalee 14-15, D-53115 Bonn, Germany 
}\\
$^{5}$ \quad {Tartu Observatory, University of Tartu, Observatooriumi 1, Toravere, 61602, Estonia; colin.folsom@ut.ee (C.P.F.)} 
}

\corres{Correspondence: kor@sirrah.troja.mff.cuni.cz}

\abstract{
We summarise the properties and nature of a peculiar group of B-type stars called FS CMa stars.
These stars show the B[e] phenomenon, i.e., their spectra 
exhibit both 
forbidden emission lines and infrared excess. 
Such properties point to 
an extended circumstellar gas and dust component. Although the phenomenon has been explained 
in most B[e] stars, the origin and nature of FS CMa stars is 
disputed.
Here, we focus on the merger hypothesis, for which
evidence has 
recently
been discovered.
}


\usepackage{natbib}         
\usepackage{xcolor}
\definecolor{darkred}{rgb}{0.545,0.0,0.0}

\newcommand*\aap{A\&A}

\newcommand*\aaps{A\&AS}

\newcommand*\apj{ApJ}
\newcommand*\apjl{ApJ}

\newcommand*\apjs{ApJS}

\newcommand*\memsai{Mem.~Soc.~Astron.~Italiana}
\newcommand*\mnras{MNRAS}

\newcommand*\pasp{PASP}

\newcommand*\ssr{Space~Sci.~Rev.}

\newcommand{\cervene}[1]{\color{darkred} {\bf #1} \color{black}}

\DeclareRobustCommand{\ion}[2]{\textup{#1\,\textsc{\lowercase{#2}}}}
\newcommand*\element[1][]{%
  \def\aa@element@tr{#1}%
  \aa@element
}

\newcommand{\kms}{km~s$^{-1}$}

\newcommand{\mlr}{$\dot{M}$}

\begin{document}


\section{Introduction}

B[e] stars are B-type stars with forbidden emission lines and infrared excess. The name 
was introduced by Conti at IAU Symposium 70 (1975). Lamers et al. \cite{Lamers98} found stars with these properties
among supergiants (sgB[e] stars), pre-main sequence Herbig Ae/Be stars (HAeB[e] stars), compact planetary nebulae
(cPNB[e]), and symbiotic stars (SymB[e] stars). For more details about these interesting objects, we refer to several very
good reviews 
\cite{Zickgraf98,Zickgraf00,Swings06,Miroshnichenko06_zavorky,Hillier06,Oudmaijer17}. 

Here, we focus on a fifth group of B[e] stars introduced later by Miroshnichenko \cite{Miroshnichenko_FS_CMa}.
He noticed that the B[e] stars that remained unclassified have similar properties. 
These stars do not fit into any of the four original groups, or may be in more 
than one of
them. However, some of them were 
not classified 
because of a~lack of good data. This is why the list of FS~CMa stars also contains  
some symbiotics, supergiants of lower luminosity, young planetary nebulae, or Be/X-ray binaries.


\section{Observed Properties}

Despite the uncertain classification, their spectral properties are unique. Based on new data,
we were able to specify the following properties and behaviour:

\subsection{Forbidden 
 Emission Lines}

The presence of forbidden emission lines of neutral and singly ionized atoms is one of the criteria 
defining the FS~CMa group \cite{Miroshnichenko_FS_CMa}. The oxygen doublet 
[\ion{O}{i}] $\lambda \lambda$ 6300, 6364~\AA \, is always present. Especially important are lines of
[\ion{S}{ii}]  $\lambda \lambda$ 6716, 6731~\AA\, and [\ion{N}{ii}]  $\lambda \lambda$ 6548, 
6583~\AA\, 
that allow the usage of nebular diagnostics, e.g., 
Baldwin, Phillips \& Terlevich (BPT) 
diagrams \cite{BPT}. 
Forbidden emission lines are very narrow (FWHM $\approx$ 50 \kms), symmetric, and sometimes double-peaked. 
They are used as tracers of the conditions in the disk. However, this has to be carried out with caution,
since stimulated emission plays a key \linebreak role here.

\subsection{Dust Emission}

Sheikina at al. \cite{Sheikina_warm_dust_00} found that IR excess of some 
unclassified B[e] stars steeply decreases longward of 25 {$\mu$}
m. Later, Miroshnichenko \cite{Miroshnichenko_FS_CMa} 
used this property as one of the key characteristics of FS~CMa stars. The episodes of dust creation have been 
observed in some stars, e.g., in FS~CMa~\cite{Swings71, deWinter97}, MWC~349~\cite{Josselin98} 
and MWC~342~\cite{Miroshnichenko99_MWC342}. The dust region is very clumpy. Sometimes, huge clouds 
form and may cover a star, strongly affecting the photometry. Such events have been observed in, e.g., FS~CMa~\cite{deWinter97}.

Miroshnichenko \cite{Miroshnichenko10_Mexico} detected a weak silicate emission in most FS~CMa stars.
However, the dusty disk has a more complicated structure. Gauba et al.~\cite{Gauba03} found that the inner part 
of the disk around MWC~939 is composed of graphite grains, while the outer part is composed of silicates. 
Even more complex is the dust disk around HD~50138 \cite{Varga19}. Close to the star, the disk is composed 
of forsterite.  In farther regions, there is forsterite  mixed with enstatite and the most distant 
parts contain amorphous silicates.

\subsection{UV Radiation and Resonance Lines}

UV radiation plays a~key role in the spectrum formation of FS~CMa stars and affects 
their entire spectrum and its variability. The UV radiation passing through the circumstellar matter 
is strongly absorbed in spectral lines of iron group elements, creating 
the so-called iron curtain. The absorption may be strong enough to reduce the UV flux 
by about an order of magnitude compared to a classical Be star of the same temperature \cite{Ivan-Pariz}.
Since the density in the circumstellar envelope is sufficiently low to enable the
non-local thermodynamic equilibrium 
regime, this absorbed energy is partially responsible for the emission of spectral lines of these species in the visible and IR regions.
The redistribution of the energy in the circumstellar matter is also detectable  in 
the 
Johnson U-band
photometric filter 
\cite{Bergner90}. This way, it is very easy to study its variability, e.g., in MWC~342~\cite{Bergner90}.

The circumstellar matter is not static and it may be the case that, due to the Doppler shift, 
a~photon from a spectral line of one element is absorbed in a spectral line of another element. This effect is especially 
strong in resonance lines because they have a very high transition probability. The temperature and
velocity of the matter around FS~CMa stars arrange the coincidence of L$\beta$ and resonance line
of neutral oxygen 1026~\AA. This way, the variability of the Balmer lines and forbidden oxygen lines
is coupled, even if their line-formation region is different. 

Almost every resonance line is in the UV region. However, \ion{Ca}{ii} H and K lines 
in the near-UV region
are
observable with ground-based telescopes and \ion{Na}{i} D1 and D2 fall into the visible region. 
This allows the study of their variability. Most FS CMa stars show symmetric emission of these lines; 
nevertheless, asymmetric emission, P~Cygni profiles, inverse P~Cygni profiles as well as rarely observed 
absorption (distinct from interstellar) have also been detected. 
The absorption, if detected, is very variable. The pure emission lines are stable on time scales of weeks
\cite{Pogodin97, deWinter97, Miroshnichenko02_388, Marchiano12}; however, they do show changes over 
a~longer period of time \cite{hab}. The resonance lines of \ion{Ca}{ii} and \ion{Na}{i}  occasionally show  discrete absorption
components. They were reported by Babcock~\cite{Babcock58APJS} for FS~CMa and 
Pogodin~\cite{Pogodin97} for HD~50138. We also detected these discrete absorption lines in MWC~342, 
AS~225, AS~174, HD~328990, and HD~50138 \cite{hab}.

\subsection{Hydrogen Lines }

FS CMa stars are 
typified
by the strong emission of the Balmer lines. For example, the equivalent width
of the H$\alpha$ line may reach 
even $-400$~\AA. 
The higher the member of the Balmer series, the weaker and narrower the line is. However, this is just the radiative transfer effect that 
 every emission line star shows. More important is the large number of Balmer and Paschen lines. 
One may usually count around forty members. They set the limits of the density and turbulent velocity 
using the Inglis--Teller formula \cite{Inglis_Teller}. The density and also turbulent velocity is very low 
in these outer parts, e.g., for FS~CMa are the limits of the ion density 
$\approx$ $1\cdot 10^{11}$ cm$^{-3}$ and turbulent velocity $v_{\rm{turb}}<40$ km/s \cite{Radek_Praha}. 

The H$\alpha$ line is usually double-peaked with 
a~weaker
blue component. To date, the blue peak has been found
to be more 
intense
than the red one only once in HD~50138 \cite{Terka16}. H$\alpha$ lines of lower intensities
may show weak absorption wings. The line profile changes on scales longer than a few days. However, very frequently, humps that move across the line \linebreak are detected.

\subsection{Molecules}

Although the effective  temperatures of the central stars are too high to allow the presence of molecules, the inner parts
of the disk serve as an effective shield against the stellar UV radiation and prevent the dissociation of molecules. 
The most frequently detected molecule is CO. That is natural, because this molecule has high binding energy.
Besides CO, SiO, TiO, H$^{+}_{3}$, and PAHs were also found. For a~recent summary of detected molecules, we refer to 
table A.5 
of
Kor{\v{c}}{\'a}kov{\'a} \cite{hab}.

\subsection{Magnetic Field}

A~magnetic field has been detected only in one FS~CMa star, namely 
IRAS 17449+2320 \cite{IRAS_ja_mag}. 
This star is also the first magnetic B[e] star. The magnetic field strength is comparable 
to that in the strongest magnetic Ap stars. The values of
the
 magnetic field modulus (around 5~kG) 
as well as longitudinal magnetic field strength vary with the rotation period of 36.12 days \cite{Iris_iras}.

\subsection{Variability}

The spectral lines vary on many time scales. Sometimes it is even better to talk about irregular behaviour. 
The absorption lines of helium and metal lines show night-to-night variations. Their line profile may 
show extreme changes from pure absorption to pure emission, through the P Cygni profile as well as inverse P Cygni profile,
or absorption with emission wings. The H$\alpha$ line and pure emission lines of metals show multiperiodic
behaviour on scales of weeks and months. The behaviour of the H$\alpha$ and \ion{He}{i} 
5876~\AA \, 
lines is described very well by Pogodin \cite{Pogodin97} for HD~50138. Forbidden emission lines are the most stable,
but they also change on scales of months and years \cite{Honza12, Blanka, MWC728, Terka16}. 
To such a complex variability contribute temperature,  
density, and motion in the atmosphere and the disk, and also
non-LTE effects, multiple scattering, and coincidence of wavelengths of resonance lines in the circumstellar gas. 
In other words, strong absorption in the UV and variable UV continuum affect the spectral lines in 
the visible and the IR. 

Along with the spectral lines, the continuum varies as well \cite{Blanka}.
As a~consequence, the intensity of all spectral lines decreases or increases. Depending on the continuum source,
the red and blue part of the spectra may behave differently \cite{IRAS_ja_mag}. 
The changes 
of the continuum are caused mainly by obscuring of the dust with different densities
and \linebreak dust reflection.

\subsection{Stellar Masses}

The 
 stellar masses for individual stars are determined based on
stellar evolutionary tracks in the 
Hertzsprung--Russell diagram (HRD). The obtained value is affected by the uncertainty of stellar parameters and 
assumptions used for the calculation of the evolutionary tracks. Both these effects are very large in the case of FS CMa stars.
Due to the circumstellar matter, the standard methods of the determination of effective temperature usually fails. 
Sometimes the guess based on the presence of 
different ionisation stages of individual elements provides an even more
accurate value than the fit by 
synthetic spectra. Nevertheless, Miroshnichenko et al. \cite{Miroshnichenko01} obtained
the first HRD.
The second problem lies in the choice of 
evolutionary 
tracks.
Among FS~CMa stars, 
some strongly interacting binaries, or even mergers, are hidden.
For mergers, only a few evolutionary models exist.

Dynamical masses from binary motion provide a more accurate method of mass determination.
Barsukova et al. \cite{Barsukova23} derived the 
mass of the primary between 10 and 12~M$_{\odot}$ and a~secondary between 0.98 and 1.77 M$_{\odot}$ for CI Cam based on the orbital solution 
with the inclination angle determined from interferometry \cite{Thureau09}. However, the question is whether CI~Cam is a typical 
member of the FS CMa group, since it has been classified as Be/X-ray binary \cite{Hynes02, Robinson02} or B[e] supergiant \cite{Bartlett19}. 
Miroshnichenko et al. \cite{MWC728} determined the mass function for MWC 728 ($2.3 \times 10^{-2}$). 
A~more exotic system is AS~386, where the primary star has mass of around 7~M$_{\odot}$ and 
the secondary is very likely a black hole \cite{Khokhlov18}. 
The companions with similar masses between 6 and 8~M$_{\odot}$ has a system HD~327083 \cite{Nodyarov22}.
The mass of the primary of MWC 645 was determined to be around 7~M$_{\odot}$ and its secondary around 2.8~M$_{\odot}$~\cite{Nodyarov_MWC645}.
Recently, the masses of the system were found for the most brightest FS~CMa star, 3 Pup.  
The primary star has a mass around 8~M$_{\odot}$ and the secondary 0.75~M$_{\odot}$~\cite{Miroshnichenko20_3_Pup}. For more details, we refer to 
Miroshnichenko et al. \cite{Miroshnichenko23_binaries}, who present 
an HRD that also includes some questionable binary systems.


\section{Nature of FS CMa Stars}

The precise determination of parameters of the central stars is difficult because of the surrounding gas and dust.
We do not have reliable information for some of these stars. Therefore, the usage of standard 
techniques is limited and many observations may be easily misinterpreted. 

\subsection{Herbig Ae/Be Stars}

Spectra of FS CMa stars are almost identical to Herbig Ae/Be stars
(for 
 recent review of Herbig Ae/Be stars see \cite{Brittain23}).
However, the FS~CMa stars are far from
star forming regions \cite{Miroshnichenko_FS_CMa}, 
a~key element of HAeBe classification.

\subsection{Post-AGB Stars}

FS CMa spectra are also similar to post-AGB stars. Several FS CMa stars are in the list of post-AGB 
stars, e.g., Hen3-938, Hen3-847, or MWC~939. 
Miroshnichenko 
 \cite{Miroshnichenko18} 
summarized
the properties of FS CMa stars and found the differences
between these two groups. Post-AGB stars with masses larger than 5 M$_{\odot}$ have the peak of the IR excess at wavelengths 
longer than 30~$\mu$m, while FS CMa stars from 10 to 30~$\mu$m, i.e., the dust is warmer and hence closer to a~star in FS~CMa objects. 
Moreover, massive post-AGB stars evolve so quickly that the temperature changes 
should be observed on decadal timescales.
Post-AGB stars of lower masses, around 1~M$_{\odot}$, show weak emission lines. However, this is the case
of a single star, as already mentioned by Miroshnichenko \cite{Miroshnichenko18}. 
Post-AGB binaries also have dust close to the star. Its emission peak is around 10~$\mu$m  \cite{Van_Winckel18}. 
However, almost all post-AGB binaries have effective temperatures that are too low to be B- or early A-type stars \cite{Kluska22}.
Nevertheless, increasing number of observations and more precise models may lead to re-classification of some 
objects from these two groups.


\subsection{Classical Be Stars}

Even if it may seem that FS CMa stars are only an extreme case of classical Be stars, there are significant 
differences. Classical Be stars 
(recent 
 review \cite{Rivi_Robert})
do not have dust surrounding them, and if they do, it is only a small amount. 
Moreover, classical Be stars are fast rotators, while FS~CMa stars are slow rotators
(\cite{hab}, Table A.7). 
Therefore, the mechanism of the creation of their gaseous disk must be different. 

\subsection{Binarity}

Binarity is currently 
an often-discussed 
scenario \cite{Miroshnichenko15_dvojhvezdy, Miroshnichenko_binaries20}.  
It provides the simplest explanation of the large amount of circumstellar gas and dust. Dust is particularly problematic,
since it needs special conditions for its creation. 
A detailed discussion of dust formation is provided by Kor\v{c}\'{a}kov\'{a} \cite{hab}.
Here, we briefly summarise the important points in favour of and against the current binarity hypothesis. \vspace*{2mm}

\subsubsection{Mass-Loss Rate (\mlr)}

The large amount of circumstellar matter is the strongest argument for the binarity of all FS~CMa stars.  
\mlr\, has been calculated only for HD~87643 \cite{Pacheco82}, AS~78 \cite{Miroshnichenko_AS_78}, and IRAS 00470+6429 \cite{Carciofi10}.
The derived values of \mlr\ from $2.5 \cdot 10^{-7}$ to $1.5 \cdot 10^{-6} M_{\odot} \cdot \rm{yr}^{-1}$, 
are not possible to  reach 
by
 the radiatively driven wind of a main-sequence star of such a~mass. 
All three models assumed a $\beta$ velocity law or a similar one, i.e., 
monotonously increasing velocity until the limiting value of the terminal velocity. 
However,  an expanding decelerating layer in MWC~342 \cite{Blanka} was later detected.
The presence of moving layers was explained by MHD simulations of Moranchel Basurto et al. \cite{AliciaI, AliciaII}. 
If the matter is accumulated around the star, the approximation of the freely expanding matter gives 
a~significant overestimate of  \mlr. \vspace*{2mm}

\subsubsection{Detection of the Secondary}

The detection of a companion has been reported in many FS~CMa stars. 
The binary hypothesis is supported by the fact that the fraction of binaries among early B-type stars is about 95\% \cite{Villasenor21}.
However, this ratio is only about 30\% for late-type main-sequence stars \cite{Vanbeveren17} and 
this is the temperature range of most FS CMa stars.
Indeed, 
evidence of a secondary star has been found only in about one-third of FS~CMa stars \cite{Miroshnichenko15_dvojhvezdy}.
The detection of a~secondary star may be explained by another phenomenon, low-quality data, or misinterpretation of the observations
in many cases. 
For example, spectroastrometric observations indicate  a secondary in HD~50138, FS~CMa and HD~85567 \cite{Baines06}.
However, the spectroastrometric signal is sensitive to any deviation from spherical symmetry \cite{Whelan08, Blanco14}. 
The non-uniform disk discovered interferometrically in HD~50138 \cite{Kluska16} mimics a companion in this case.
Unambiguous evidence for a~secondary would be provided by radial velocity measurements.
However, sufficient data are only available for
MWC~623 \cite{Honza12},
MWC~342 \cite{Blanka},
MWC~728 \cite{MWC728},
HD~50138 \cite{Terka16},
AS 386~\cite{Khokhlov18}, 
IRAS~07080+0605 \cite{Khokhlov22},
and
IRAS~17449+2320 \cite{IRAS_ja_mag}. 
Regular periodicity has been found only in AS~386 and MWC~728. 
Other stars (in addition to those discussed here) also show radial velocity changes.
However, those variations are generally quasiperiodic and are not connected with orbital motion, 
or it is not possible to reject the quasiperiodicity based on several measurements \cite{hab}. \vspace{2mm}

\subsubsection{Central Quasiemission Peak}

A small emission peak at the system velocity sometimes appears in the H$\alpha$ line, 
similar to the classical Be stars.
This emission is taken as further support of the binarity, because it is assumed that it originates near the Lagrangian point L1 \cite{Khokhlov22}. 
However, Hanuschik~\cite{Hanuschik95} showed that it is only a radiative transfer effect. The emission is formed 
through the circumstellar disk that is optically thick in the line and optically thin in the continuum. 
In the case of FS~CMa stars, there is an even simpler explanation: a moving hump revealing inhomogenities in the disk 
is detected accidentally at zero velocity. \vspace{2mm}

\subsubsection{Spectrum of a Hot and Cold Component}

The spectrum of MWC 623 is a composition of B4III and K2II (or K2Ib) spectral type~\cite{Zickgraf01}. 
This is a definite proof of it being a binary 
system, however in ordinary stars. 
FS~CMa 
stars are surrounded by geometrically and optically thick disks.
The cold component spectrum may be formed through the disk. This explanation is supported by 
a~modelling
of variability of bisectors of the H$\alpha$ line in MWC~623 \cite{2028_Cyg_bisector}. The orbital motion 
alone is not able to describe the changes. The radial velocities support this idea \cite{Honza12}.
The lithium resonance doublet, which is the strongest in MWC~623, 
is
likely formed in a similar manner. \vspace{2mm}

\subsubsection{Lithium Lines}

About half of FS~CMa stars show the lithium resonance doublet at 
6708~\AA \,
\cite{lithium_conference}. This line should not be observed in hot B-type stars because lithium is ionised in their hot
atmospheres (the ionization potential of \ion{Li}{i} is about 5.4 eV). Therefore, the presence
of \ion{Li}{i} lines has been taken as the evidence of a secondary. However, FS~CMa stars have
an extended circumstellar region. The lower-temperature circumstellar gas causes the additional absorption. 
In support of this idea is the appearance of a sharp absorption in \ion{He}{i} line 
5876~\AA \,
in stars with \ion{Li}{i} lines \cite{lithium_conference}. 
This sharp absorption overlapping the emission is likely created through the circumstellar disk. 
There are no radiative transfer calculations of these objects that are sufficiently accurate to confirm 
this guess. However, a
similar behaviour is also shown by some Herbig Ae/Be stars. Based on 
polarimetric measurements, Alecian et al. \cite{Alecian13} 
noticed that lithium lines in these Herbig Ae/Be stars do not originate in the atmosphere of the secondary. \vspace{2mm}

As we can see, 
 many effects and phenomena play here   
an important role and it is necessary to follow the original papers with the observations.

\subsection{Mergers}

There is a group of FS CMa objects whose observed properties cannot be explained by known stellar types. 
Based on the latest observations and numerical simulations,
a merger scenario becomes  very promising. We summarize the points in favour of a~merger idea in the following section.


\section{Post-Mergers Among FS CMa Stars}

First, it is necessary to note here that the post-merger origin is not an explanation
for every FS~CMa star. The definition of FS~CMa stars (see the introduction) is not 
a~sufficiently sharp criterion and the current list of FS~CMa stars contains some interacting binaries, young 
planetary nebulae, symbiotic stars, Herbig Ae/Be stars, post-AGB stars, as well as supergiants. 

The idea of mergers among FS CMa stars is relatively old. However, it has not been discussed in the literature
because of the lack of evidence. 
The first who opened the discussion was de la Fuente et al. [68]. 
They discovered two FS CMa stars in the central parts of two clusters. They explained this observation 
as the consequence of the capture of two stars in dense regions. However, as shown by
Dvo\v{r}\'{a}kov\'{a} et al.  \cite{Nela}, the mergers may be found with almost
the same probability in the central as well as outer parts of clusters.
Still, there is no strong and straightforward evidence, 
but rather a combination of many observations and simulations that all come together to create 
a consistent picture:

\subsection{Magnetic Field}

The strength of the magnetic field found in 
IRAS 17449+2320 
is comparable with the strongest magnetic Ap stars. 
Such a strong magnetic field may be generated by the mixing during the merger process \cite{Schneider20}.
Moreover, the spectropolarimetry of 
IRAS 17449+2320 
shows that the longitudinal magnetic field strength and magnetic modulus changes with a~period which corresponds 
to the rotation period rather than the orbital one \cite{Iris_iras}.

\subsection{Slow Rotators}

The magnetic field generated during the merger and the later outflow in the form of a~wind
slows down the newly born star very effectively, e.g., \cite{Schneider20}.
Indeed, FS~CMa stars are rather slow rotators  
\cite[][table A.7]{hab}.
Some of the published values of the rotation velocity are very likely overestimated.
This is the case for stars with  a~magnetic field that causes the line split or 
broadening for less sensitive lines or a weak field. To obtain a real value of $v \sin i$, 
synthetic spectra with the magnetic field must be used.

\subsection{Space Velocities}

Most FS CMa stars have a~large space velocity \cite{Nela}. Their peculiar W component is especially important for us.
It indicates that the star was very likely kicked from the birth cluster due to 
an accidental interaction with another star in the cluster. If it is
a binary, it usually has a short semi-major axis and a large eccentricity. 
Consequently, the tidal forces circularise the orbit. As a consequence of the momentum conservation, 
the semi-major axis shrinks and binary may merge. Let us highlight at this point 
that binary components may merge not only in triple systems, or during the common envelope phase, 
but it may also be a natural consequence of the dynamical evolution of the 
birth cluster. Dynamical processes may also lead to mergers in clusters \cite{Nela}, 
i.e., the large space velocity is in favour of a merger; however, a small space velocity does not reject the merger origin. 

\subsection{Li Overabundance}

Despite the high temperature of B-type stars, about half of FS CMa stars show lithium lines (see the previous section).
Lithium resonance doublet is very strong in some of these stars, which
points to a~lithium overabundance in these objects. Lithium overabundace
has been found in about half of red novae \cite{Kaminski23}, a nova-like event which occurs during a~merger.

\subsection{Hertzsprung--Russell Diagram (HRD)}

Most of FS CMa stars are located around the terminal-age main sequence in the HRD, see,
e.g., \cite{Miroshnichenko_HRD}. According to  Schneider et al. calculations \cite{Schneider20} 
of the evolution of a~merger, the newly born star 
becomes a slow rotator around the terminal-age main sequence (TAMS). 

\subsection{N-Body Simulations}

\begin{figure}[H]
\includegraphics[width=0.9\linewidth]{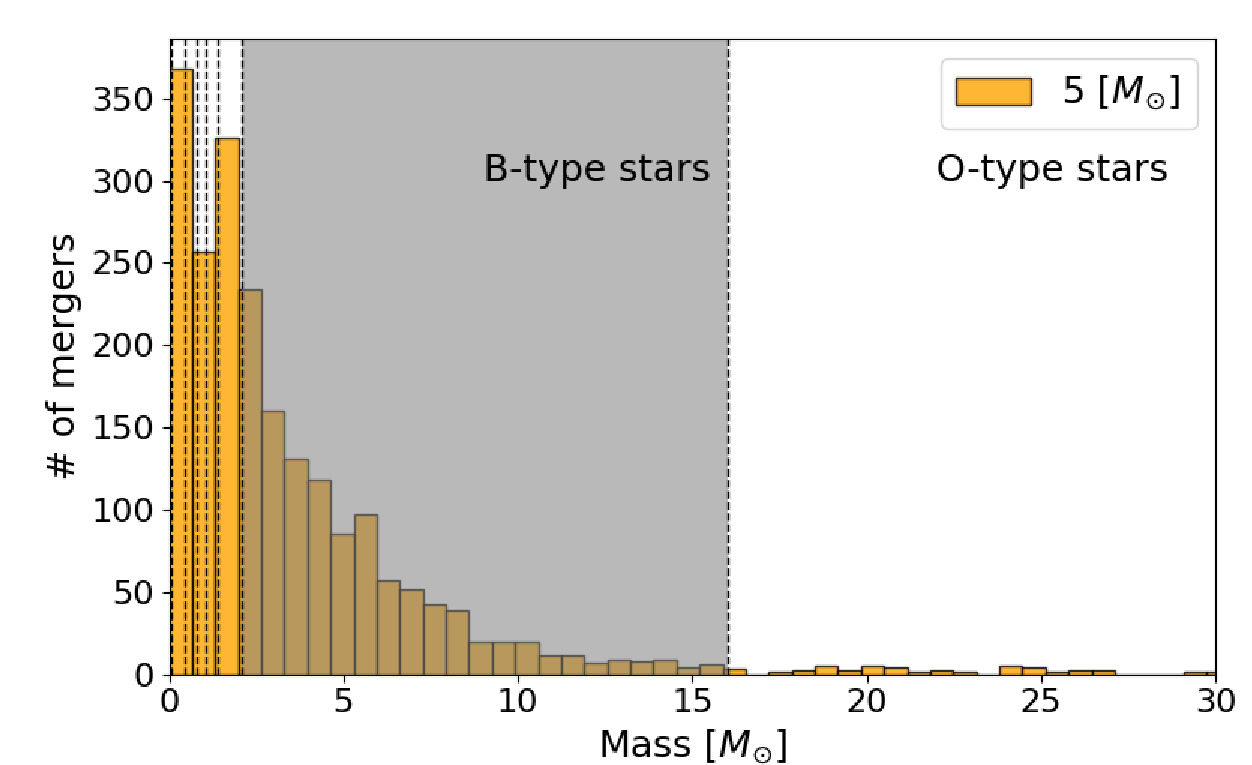}
\caption{Histogram of mergers according to spectral types
from Dvo\v{r}\'{a}kov\'{a} et al. \cite{Nela}.
 \label{Nbodyfig}}
\end{figure}

The N-body simulations of the dynamical evolution of 510 clusters analysed by
Dvo\v{r}\'{a}kov\'{a} et al. \cite{Nela} show that about 1.7\% of all
stars are involved in mergers. About 50\% of all mergers are of the spectral type B. This
is a natural consequence of the initial mass function, initial distribution of the binary parameters and 
large range of masses of B-type stars (see Figure~\ref{Nbodyfig}). 
We also provide a more detailed Table \ref{Nbodytab}.  
The table as well as the figure show the results of simulations
over Hubble time. To obtain the current observed ratio of mergers in the sky, it is necessary to include 
the lifetime of individual spectral types and galactic synthesis (Dvo\v{r}\'{a}kov\'{a} et al., in preparation).
Note 
that these simulations very likely underestimated the number of mergers, because the dynamical evolution was 
calculated only for single stars and binaries. More hierarchical systems may also lead to the merger. 
The detailed description of the evolution of the triple systems 
is provided by Toonen et al.~\cite{Toonen22}.
\vspace{-5pt}
\vspace{-5pt}

\begin{table}[H] 
\caption{The ratio of mergers of individual spectral types
from Dvo\v{r}\'{a}kov\'{a} et al. \cite{Nela}.
\label{Nbodytab}}
\begin{tabularx}{\textwidth}{CCCCC}
\toprule
\textbf{Spectral} & \textbf{Stars at the}     & \textbf{Stars Involved}  & \textbf{Merger}         & \textbf{Merger} \\
\textbf{Type}     &  \textbf{Beginning [\%] } & \textbf{in Mergers [\%]} & \textbf{Products [\%] } & \textbf{Ratio [\%]} \\
\midrule
O        &    0.12          &   3.32         &  3.46          & 26.22  \\
\textbf{B} &  \textbf{3.53} & \textbf{48.15} & \cervene{50.48}  & \cervene{12.54} \\   
A        &    2.77          &   11.49        &  15.18         & 4.80 \\
F        &    3.28          &   8.95         &  7.11          & 1.90 \\
G        &    4.05          &   6.53         &  3.88          & 0.84 \\
K        &    15.29         &   10.71        &  10.29         & 0.59 \\
M        &    70.97         &   10.82        &  8.81          & 0.11 \\
\bottomrule
\end{tabularx}\vspace*{1mm}
\noindent{\footnotesize{
Let us follow the dynamical evolution of stars of the spectral type B. According to the initial mass function, 
about 3.53\% of all stars are of the spectral type B at the beginning of the simulation. 48.15\% of stars used for the merger creation are 
of the spectral type B. This causes that 50.48\% of all merger products become finally B-type stars (highlighted in red). This also means a~high percentage of mergers among B-type stars (12.54\%).
Note, that the highest fraction of mergers among the given spectral type is in O stars. However, there are not so many O-type stars; 
therefore, the highest chance to find a post-merger is among B-type stars.                                                                                                                          
}}
\end{table}

\subsection{MHD Simulations}

\begin{figure}[H]

\includegraphics[width=0.49\linewidth]{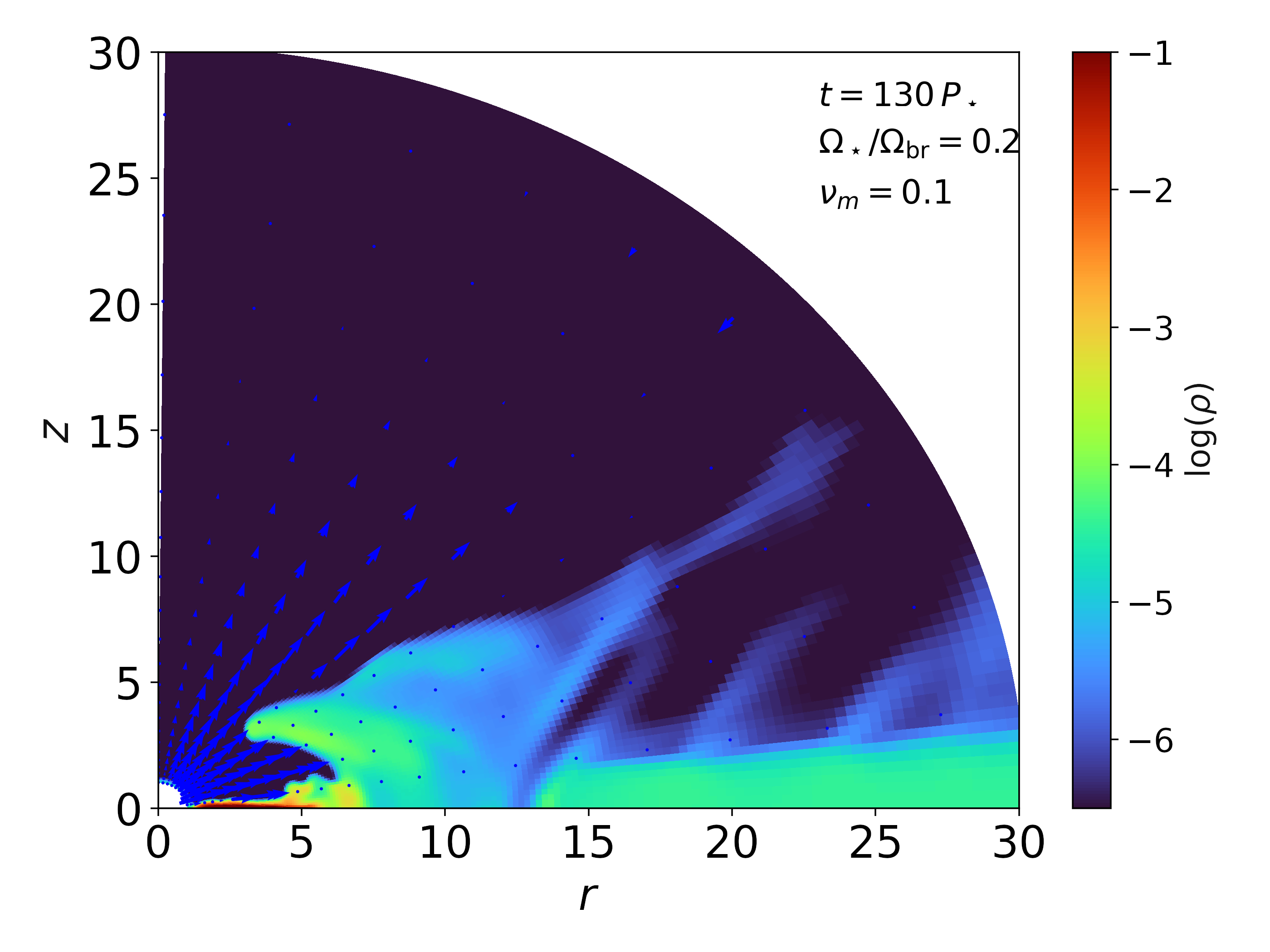}
\includegraphics[width=0.49\linewidth]{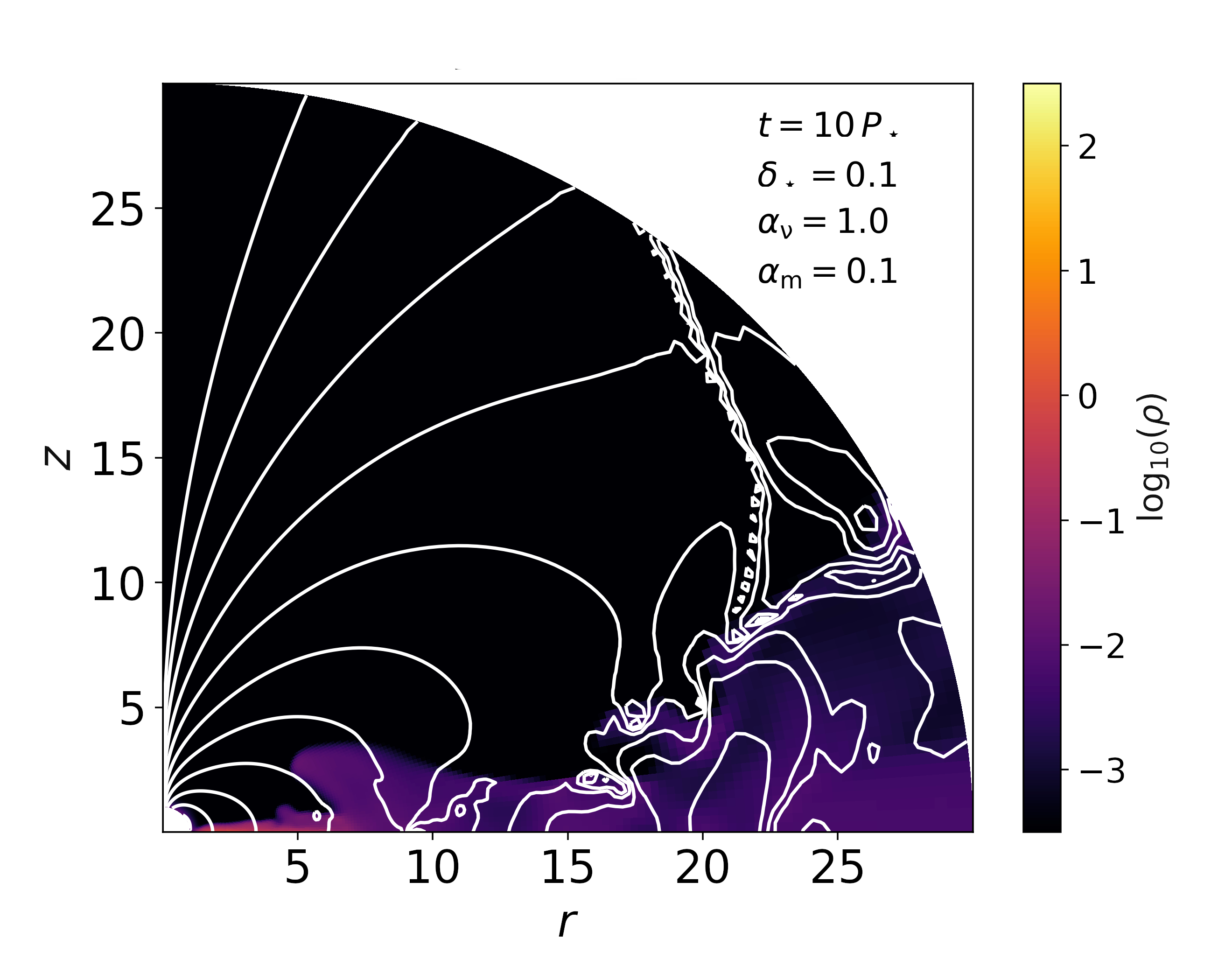} \\
\includegraphics[width=1.0\linewidth]{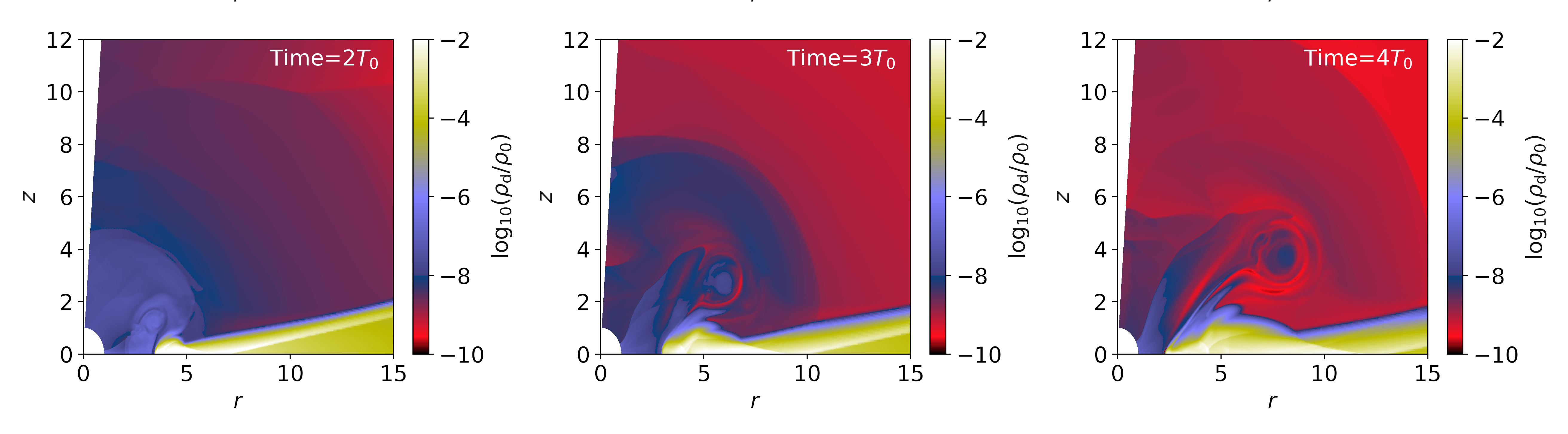} 
\caption{2.5D MHD simulations of a FS CMa post-merger. 
\textbf{upper panel}: The disk structure and accretion channels. The magnetic field
lines are indicated by white lines
(based on the work of Moranchel-Basurto et al. \cite{AliciaII}).
\textbf{bottom panel}: The evolution of a hot plasmoid (based on Moranchel-Basurto et al. \cite{AliciaI}).
\label{MHDfigs}}
\end{figure}

Moranchel Basurto et al. \cite{AliciaI, AliciaII} performed the first 2.5D MHD simulations 
of a~post-merger based on parameters determined from the observations of FS~CMa stars. 
They explored different scenarios considering the low density of a corona,
sub-Keplerian disk rotating around the star, and a strongly magnetized
non-rotating and rotating star with a~dipolar configuration.
They found that due to the magnetic field,  low density of the disk and 
the relative rotation speed between the disk and the star,
the disk structure changes 
extremely
(Figure~\ref{MHDfigs}, upper panel). 
The matter concentrations and gaps are created and the disk becomes significantly thicker. 
The accretion may occur through a~thin stream in the equatorial region. It may 
also be accompanied by a~stream following the magnetic field lines, 
so-called funnel effect.
However, this funnel effect becomes unstable under some conditions. The matter may be 
captured by the magnetic field above the star. After some time, it may either be
ejected from the system or fall onto the stellar surface. The magnetic field lines become extremely 
twisted in a very short time. This configuration is highly unstable and leads to the magnetic reconnection. 
The released energy heats the matter in the corona which is fed by the wind from the disk, 
a~jet originated at the inner edge of the disk or by a~hot plasmoid (Figure~\ref{MHDfigs}, bottom panel).

These MHD simulations finally explain most of the behaviour
of FS CMa stars, including moving decelerating layers \cite{Blanka} or detection of weak Raman 
lines in some of \linebreak these stars. 

\vspace{-5pt}
\unskip

\section{Role of FS CMa Post-Mergers}

The simulations of a merger evolution by Schneider et al. \cite{Schneider20} show that 
a~star is overluminous before it reaches the main sequence. 
That points to the possibility that lower massive mergers of B spectral type  become magnetic main-sequence A-type stars. 
This transition phase offers us a~unique possibility to test models of the stellar structure and evolution.

As the star evolves on the main sequence, the strength of the magnetic field 
may decrease due to Ohmic decay and/or flux freezing. As the main-sequence lifetime is shorter for more massive stars,
the more massive FS CMa mergers 
with originally
an extremely strong magnetic field may still have a measurable magnetic field at the end of the main-sequence 
(for the effect of Ohmic decay, see observation surveys of Ap stars, \cite{Landstreet09}).
Hence, more massive magnetic FS CMa mergers may offer the explanation of a~recently 
 discovered young
strongly magnetic low-mass white dwarf \cite{Landstreet22}.

Before the merger, during it, and also after the merger, a significant amount of the matter is released,
allowing the creation of molecules and dust. Part of this material enriches the interstellar matter
(Dvo\v{r}\'{a}kov\'{a} et al., in preparation). The intergalactic medium is also slightly affected, 
because some binaries are kicked out of the birth cluster with a high speed \cite{Nela}.


\section{Concluding Remarks}

The observed properties of FS~CMa stars point to an extended region of circumstellar gas and dust 
the inner parts of which are very dynamic. 
These properties may correspond to many different classes of objects.
The list of FS~CMa stars contains interacting binaries, young planetary nebulae,
symbiotic stars, low-luminosity supergiants, or post-mergers. Up to now, only one post-merger 
has been identified: 
IRAS 17449+2320 
\cite{IRAS_ja_mag}. However, it is reasonable to assume 
that more post-mergers are hidden in this group. This scenario is supported by MHD simulations conducted by 
Moranchel Basurto et al. \cite{AliciaI, AliciaII} that explain the unusual behaviour of several FS~CMa stars, e.g., 
moving decelerating layers, Raman lines in some objects, or discrete absorption components of resonance lines. 
A~strong
support to start the systematic search of post-mergers among the FS~CMa group is provided by the
N-body simulations of Dvo\v{r}\'{a}kov\'{a} et al. \cite{Nela}. They found that a significant amount of all stars
(1.7\%) are involved in mergers and about 50\% of mergers are of  spectral type B. 

The determination of the nature of the individual objects is 
very
problematic. The observations as well as the results of the analysis may
be easily misinterpreted due to the circumstellar gas and dust. However, FS CMa post-mergers may finally explain 
long-standing issues connected to the evolution of magnetic stars. The FS~CMa stars with lower masses are potentially the progenitors of some 
magnetic main-sequence A-type stars, since  post-mergers are overluminous compared to the stars in the main sequence. Mergers also enrich the interstellar 
matter. Their contribution will likely be very small compared to those of supernovae or AGB stars; however, a~significant percentage of pre-merger binaries 
have space velocities large enough to leave the galaxy and slightly affect the intergalactic matter. Since several FS~CMa stars are strongly 
interacting binaries that exhibit dust creation, this group offers us a unique possibility to identify the whole evolutionary sequence of mergers. 

\vspace{6pt}
\authorcontributions{
spectroscopy, D.K., N.D., I.B.L., G.A.W. and C.P.F.; 
N-body simulations, N.D.
and P.K.;
spectropolarimetry,
I.B.L. and G.A.W.;
MHD simulations, A.M.B. and R.O.C.; 
photometry, O.J. All authors have read and agreed to the published version of the manuscript.
}

\funding{
P.K.
acknowledges the support of the Bonn-Prague DAAD Eastern Europe Exchange grant at Bonn University. 
G.A.W.
acknowledges Discovery Grant support from the Natural Sciences and Engineering Research Council (NSERC) of Canada.
I.B.L.
is supported by the Grant Agency of the Charles University (grant number 6124).
The work of 
R.O.C.
was supported by the Czech Science Foundation (grant 21-23067M).
Computational resources were available thanks to the Ministry of Education,
Youth and Sports of the Czech Republic through the e-INFRA CZ (ID:90254).
C.P.F.
acknowledges funding from the European Union's Horizon Europe research
and innovation programme under grant agreement No. 101079231 (EXOHOST), and
from the United Kingdom Research and Innovation (UKRI) Horizon Europe
Guarantee Scheme (grant number 10051045).
}

\institutionalreview{Not applicable
}

\informedconsent{Not applicable
}

\dataavailability{
As this article reviews a subject, no new data were created.
}

\acknowledgments{
ADS and CDS were used for preparing this document.
}

\conflictsofinterest{The authors declare no conflicts of interest.
} 



\abbreviations{Abbreviations}{
The following abbreviations are used in this manuscript:\\

\noindent 
\begin{tabular}{@{}ll}
HRD  & Hertzsprung–Russell diagram \\
TAMS & terminal-age main sequence \\
MHD  & magneto-hydrodymic \\
PAHs & polycyclic aromatic hydrocarbons \\
\end{tabular}
}

\begin{adjustwidth}{-\extralength}{0cm}

\reftitle{References}




\reftitle{References}






\PublishersNote{}
\end{adjustwidth}
\end{document}